\documentclass[%
reprint,
showkeys,
showpacs,
nofootinbib,
 amsmath,amssymb,
 aps,
]{revtex4-2}

\usepackage{graphicx}% Include figure files
\usepackage{dcolumn}% Align table columns on decimal point
\usepackage{bm}% bold math
\usepackage{siunitx}
\usepackage[colorlinks,citecolor=red,urlcolor=blue,bookmarks=false,hypertexnames=true]{hyperref} 
\usepackage{xspace}
\usepackage{textcomp}
\usepackage[gen]{eurosym}
\usepackage{verbatim}

\begin{document}

\preprint{APS/123-QED}

\title{Quantum banking system for secure cash withdrawal}% Force line breaks with \\

\author{Vineet Kumar$^{1,2}$}
\email{vineet05k@gmail.com}
%\affiliation{$^1$Center for Quantum Information and Quantum Biology (QIQB), Osaka University, 1-2 Machikaneyama, Toyonaka 560-0043,Japan}

\affiliation{$^1$Charles University, Faculty of Mathematics and Physics, Dept. of Surface and Plasma Science, Prague 8,Czech Republic}

\affiliation{$^2$Center for Quantum Engineering and Science, QuuQ, Patna 800023, Bihar, India}


\begin{abstract}
\noindent A quantum-secure withdrawal framework is presented by replacing reproducible classical credentials with quantum tokens for authentication. The protocol integrates token generation, authentication, and verification into the existing banking workflow using experimentally motivated quantum control techniques. This mitigates card cloning, credential theft, and counterfeiting.

\end{abstract}



\maketitle

%\tableofcontents

\section{Introduction}
\label{sec:Introduction}

According to the Reserve Bank of India (RBI), India had approximately 250,000 automated teller machines (ATMs) at the end of financial year 2025 (as of March end). Public sector banks accounted for \(53.19\,\%\) of the total ATM network, followed by private sector banks (\(30.72\,\%\)), white label ATM (WLA) operators (\(14.43\,\%\)), and other banks, including foreign and small finance banks (\(1.67\,\%\)) \cite{ET_ATM_FY25_2025,TOI_ATM_FY25_2025}. Globally, approximately 2.9 million ATMs were in operation in 2024 \cite{DatosInsights_ATM_2024}. Although the global ATM network has declined in recent years, the scale of this infrastructure underscores the importance of cash-based transactions and highlights the widespread reliance on ATM services for everyday financial activities, despite the rapid expansion of digital payment systems.

With the widespread deployment of ATMs, the network remains susceptible to authentication-related financial crimes that exploit the copyable nature of classical credentials, enabling adversaries to duplicate card information, steal credentials, and counterfeit user credentials. Motivated by the principles of quantum money and the quantum no-cloning theorem \cite{wootters1982single,wiesner1983conjugate,bozzio2018experimental}, the presented quantum-secured cash withdrawal framework employs on-demand, non-copyable quantum tokens for authentication, thereby providing a secure transaction. 

To realize this framework, a Quantum Teller Machine (QTM) ecosystem comprising a QTM card, which provides the physical platform for the realization of quantum tokens, and a QTM terminal, which performs token generation, authentication, and transaction, is considered. Its deployment requires coordinated enhancements to both the ATM hardware and the banking back-end. The purpose of this paper is to present the design of a QTM ecosystem and its underlying quantum authentication protocol. To demonstrate its practical feasibility, the framework is implemented using a solid-state spin-defect platform, with the nitrogen-vacancy (NV) center adopted as a representative physical realization. 

The structure of this paper is as follows. Section \ref{sec:ATM Withdrawal} reviews the conventional ATM withdrawal process and its security limitations. Section \ref{sec:QTM Withdrawal} presents the proposed QTM-based withdrawal protocol. Section \ref{sec:Physical Realization} discusses the physical realization of the protocol using NV-center spin qubits. Finally, Section \ref{sec:Conclusions and outlook} concludes the paper.

\section{ATM Withdrawal} 
\label{sec:ATM Withdrawal}

In general, an ATM-based cash withdrawal process consists of several steps to complete a single transaction, all of which rely on available classical digital data (CDD). This typically includes the card number $N_u$, the cardholder’s name $N_a$, and the card’s validity period from $T_i$ to $T_f$. Together, this information can be denoted by $\mathcal{C} = (N_u, N_a, T_i, T_f)$. The individual steps involved in the transaction are briefly outlined below for reference.

\begin{itemize}
    \item \(\textit{Card readout}\). When the card is inserted into an ATM terminal, a read operation is performed on the CDD, i.e., $\mathcal{C} \leftarrow \mathrm{Read}(\mathcal{C})$, where the operation \(\mathrm{Read}(\cdot)\) is deterministic and returns a copy of the stored data without modifying the physical state of the card, i.e., $\mathrm{Read}(\mathcal{C})=\mathcal{C}$.
    
    \item \(\textit{Credential submission}\). Following the readout, the ATM prompts the user to enter a personal identification number (PIN), denoted by \(P\), where \(P \in \{0,1,\ldots,9\}^{k}\). To protect the PIN, which is a classical string of digits, the ATM encrypts it using a standardized PIN-block formats (e.g., ISO 9564) \cite{saha2014analysis}. The ATM then transmits the card details together with the encrypted PIN to the bank through a classical communication channel, i.e., $\text{ATM} \rightarrow \text{Bank} : \{\mathcal{C'}, E(P)\}$, where \(\mathcal{C}' \subseteq \mathcal{C}\) and \(E(\cdot)\) denotes the encryption.

    \item \(\textit{Card verification}\). Upon receiving the data $({\mathcal{C}'=N_u,E(P)})$, the bank verifies the card number against its database $\mathcal{D}=\{N_u^{(1)},N_u^{(2)},\ldots,N_u^{(k)}\}$, which denotes the set of all the registered card numbers. Verification is successful only if \(N_u\in\mathcal{D}\); otherwise, the transaction is rejected. If successful, the encrypted PIN, \(E(P)\), is processed within a secure cryptographic device (SCD), such as a Hardware Security Module (HSM), where the required cryptographic operations for PIN verification are performed in accordance with ISO 9564 without exposing the PIN outside the secure cryptographic boundary \cite{PCI_PIN}.

    \item \(\textit{Transaction authorization}\). If the card and PIN verification are successful, the bank proceeds with the process by checking the current status of the card account, including $\mathcal{C}'= T_i,T_f$. If all conditions are satisfied, the bank returns an approval response to the ATM; otherwise, a decline response is returned. Upon receiving an approval response, the ATM dispenses the requested cash. 
\end{itemize}

Although ATM cash withdrawal employs secure transaction protocols, authentication still relies on CDD, making it susceptible to card cloning, credential theft, and counterfeiting. The presented quantum withdrawal mechanism replaces reproducible classical credentials with quantum tokens, thereby preventing cloning, enabling the detection of credential theft through state disturbance, and making counterfeiting both computationally and physically infeasible without the bank's secret token preparation parameters.

\section{QTM Withdrawal}
\label{sec:QTM Withdrawal}

Following consideration of the QTM card (the physical carrier of the quantum-state preparation platform) and the QTM terminal (the hardware responsible for quantum operations and measurements), the preparation of a quantum token must also be defined. Built on this infrastructure, the token preparation protocol serves as a fundamental component of the day-to-day work of the quantum banking system.

\begin{figure}
    \centering
\includegraphics[width=0.45\textwidth]{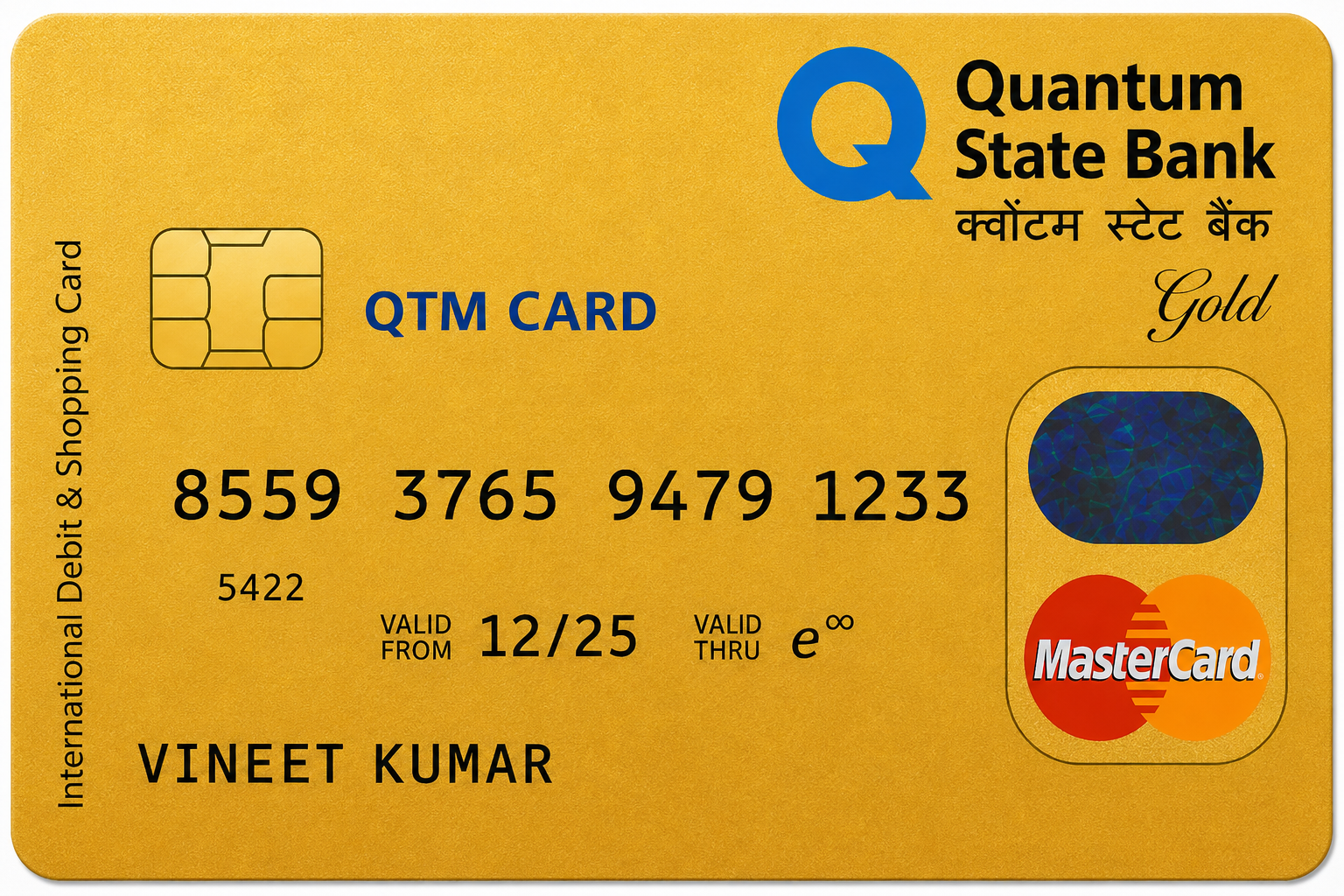}
%\vspace{0.1 em} % Adjust the negative value as needed to reduce the gap
\caption{Illustration of a conceptual QTM card incorporating an embedded quantum chip in place of a conventional ATM card chip for token management. The chip is based on a 3.0~mm $\times$ 3.0~mm DNB14 quantum-grade single-crystal diamond (Thorlabs), with a typical NV center density of 4.5~ppm, a spin dephasing time of 0.5~$\mu$s, and a spin coherence time of 10~$\mu$s. Card details are for visualization only.}
\label{fig:1}
\end{figure}

From a hardware perspective, the QTM card can be realized using solid-state spin-defect technologies capable of storing quantum states on demand, such as silicon spin devices \cite{veldhorst2015two}, silicon-carbide defect centers \cite{koehl2011room}, and nitrogen-vacancy centers in diamond \cite{childress2006coherent}. Figure~\ref{fig:1} shows the proposed QTM card.

\begin{figure}
    \centering
\includegraphics[width=0.45\textwidth]{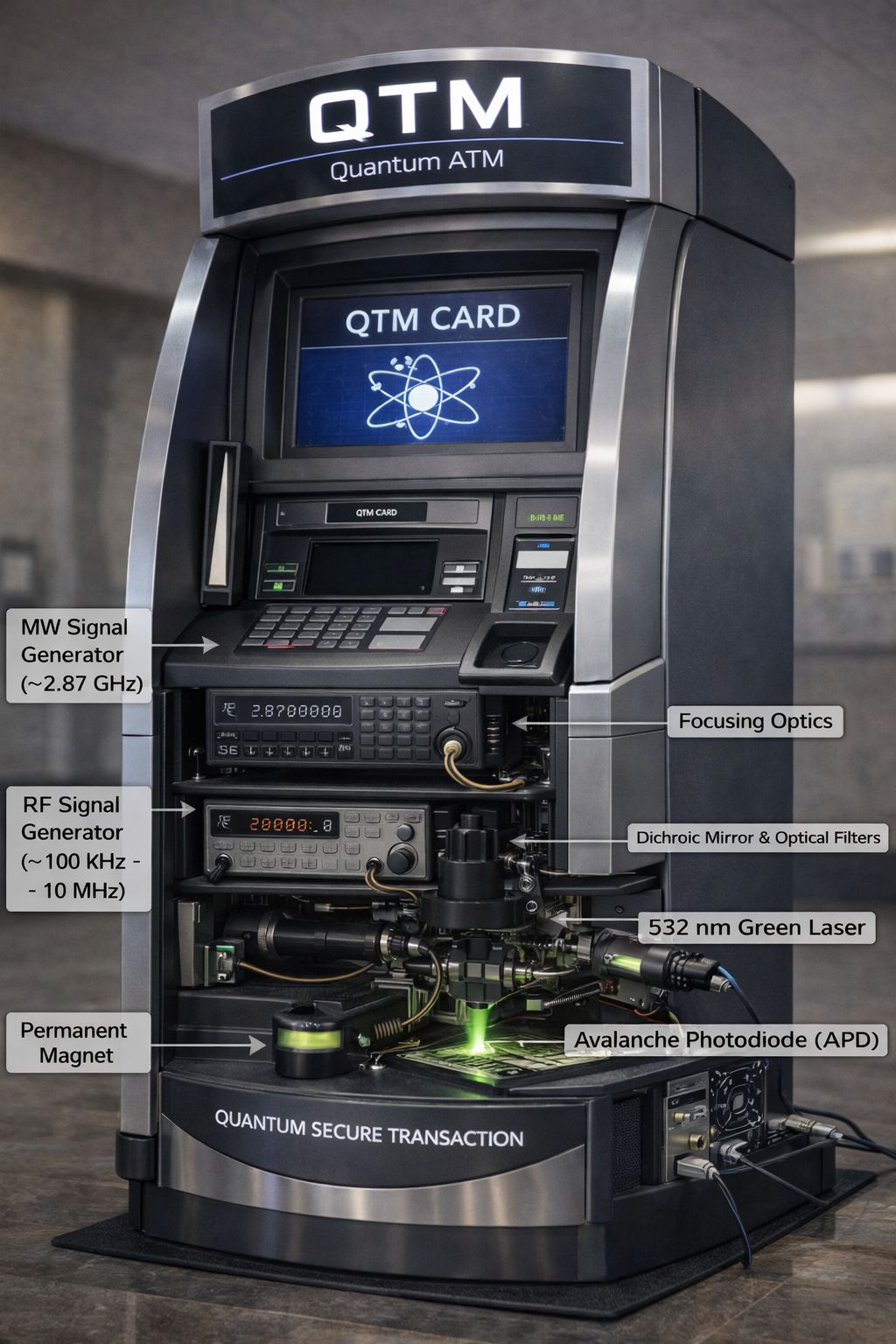}
%\vspace{0.1 em} % Adjust the negative value as needed to reduce the gap
\caption{Illustration of a conceptual QTM terminal architecture integrating several key components. Each labeled module represents a functional subsystem responsible for signal generation, optical excitation, fluorescence collection, photon detection, or synchronized system control.}
\label{fig:2}
\end{figure}

The QTM terminal, on the other hand, can be implemented as a dedicated experimental platform or rack-mounted control system that provides the state preparation, manipulation, and measurement capabilities required for token generation and authentication. In practice, a QTM terminal corresponds to a conventional ATM terminal integrated with laser sources, optical modulators, and detectors, among other supporting components. Figure~\ref{fig:2} shows the proposed QTM terminal.

With this quantum ecosystem in place, two levels of token construction can be considered. A standard quantum token employs basis randomization, whereas a hash-based quantum token additionally incorporates a cryptographic fingerprint derived from a token-specific serial number, thereby providing an additional layer of security. In both cases, the token is generated on demand by the terminal under instructions supplied by the bank, rather than being stored as a coherent quantum state within the card over the extended periods (e.g., weeks or months) typical of current ATM cards.

(i)\(\textit{Standard quantum token}\). The QTM token is considered of three classes of qubits, namely the response class (\(\mathcal{S}\)), the reference class (\(\mathcal{R}\)), and the support class (\(\mathcal{A}\)). Let \(p\), \(q\), and \(r\) denote the number of qubits in classes \(\mathcal{S}\), \(\mathcal{R}\), and \(\mathcal{A}\), respectively. The total number of qubits contained in a token is therefore \(N = p + q + r\). The raw quantum token, then can be prepared as

\vspace{-1.25em}
\begin{equation}
|\phi\rangle
=
\left(
\bigotimes_{j=1}^{p}
|u_j\rangle_{b_j}
\right)
\otimes
\left(
\bigotimes_{k=1}^{q}
|v_k\rangle
\right)
\otimes
|0\rangle^{\otimes r},
\end{equation}

\noindent where $|u_j\rangle_{b_j} = U_{b_j}|u_j\rangle$ denotes the $j^{\text{th}}$ qubit of \(\mathcal{S}\) prepared in basis $b_j\in\{X,Z\}$ with $Z$ and $X$ representing the computational basis $\{|0\rangle,|1\rangle\}$ and diagonal basis $\{|+\rangle,|-\rangle\}$, respectively, where $|\pm\rangle=\frac{|0\rangle\pm|1\rangle}{\sqrt{2}}.
$ The state $|v_k\rangle$ denotes the $k^{\text{th}}$ qubit of \(\mathcal{R}\) prepared in computational basis, while the qubit of \(\mathcal{A}\) is initialized in the state $|0\rangle$. 

The basis-selection operator $U_{b_j}$ is defined as

\vspace{-1.25em}
\begin{equation}
U_{b_j}
=
\begin{cases}
I, & b_j=Z\\
H, & b_j=X
\end{cases}
\end{equation}

\noindent where $I$ and $H$ denote the identity and Hadamard operators, respectively. The operator $U_{b_j}$ enables the preparation of response qubits on the $Z$- or $X$-basis. Thus, the random choice of $b_j$ introduces basis uncertainty for an adversary. Consequently, 

\vspace{-1.25em}
\begin{equation}
|u_j\rangle_{b_j}=
\begin{cases}
|0\rangle,\;|1\rangle, & b_j = Z,\\
|+\rangle,\;|-\rangle, & b_j = X.
\end{cases}
\end{equation}

Together with the \(\mathcal{S}\), the \(\mathcal{R}\) qubits provide supplementary resources during encoding, whereas the \(\mathcal{A}\) qubits are viewed as an operational resource. The inclusion of these different qubit classes enables flexible cascaded encoding architectures, thereby increasing the security of a token. Therefore, the standard token can be represented as $U_k|\phi\rangle$, with $U_k$ secret encoding operator together with $(p,q,r)$ and $U_b$. 

(ii)\(\textit{Hash-based quantum token}\). To elevate the protocol into a cryptographically enhanced framework, the reference class $\bigotimes_{k=1}^{q}|v_k\rangle$ can be upgraded, denoted by $|\omega\rangle$. Let $s$ represent the serial number of the token under consideration. The corresponding state is defined as

\begin{equation}
|\omega\rangle
=
\frac{1}{\sqrt{2^q}}
\sum_{x\in\{0,1\}^{q}}
e^{i2\pi H(s,x)}
|x\rangle,
\end{equation}

\noindent where $H(s,x)\in[0,1)$ is a serial dependent normalized hash function, and $x$ is the computational basis.

For the generation of \(s\), an \(n\)-bit quantum random number generator, \(\mathrm{QRNG}_n\), can be employed initially to generate a set of unique serial numbers for a given QTM card, which are stored by the bank before the card is issued to the customer. During each token preparation, a serial number \(\mathbf{s}=(s_{n-1}s_{n-2}\cdots s_1s_0)\), \(s_j\in\{0,1\}\), which corresponds to the decimal representation $s=\sum_{j=0}^{n-1}s_j\,2^j$ is selected from the set. For example, using \(\mathrm{QRNG}_{16}\), the binary string $\mathbf{s} = 1000111101100110_2$ corresponds to $s=36710$.

The phase distribution function is then given as

\begin{equation}
H(s,x)
=
\frac{
\mathrm{int}\!\left(\mathrm{SHA256}(\sum_{j=0}^{n-1}s_j\,2^j\|x)\right)
}
{2^{256}},
\end{equation}

\noindent where $\|$ denotes concatenation and $\mathrm{int}(\cdot)$ converts the hash digest into an integer. This serves as a cryptographic identity tag that uniquely binds the token to its assigned serial number.\footnote{For $q=1$ and $s=36710$, the SHA-256 hash of the string $(36710||0)$ is $\mathrm{SHA256}(36710||0) = 9e1ac4d\cdots 52a7070$, which after hexadecimal normalization yields $H(36710,0)\approx0.617596$. Similarly, $H(36710,1)\approx0.132870$. The corresponding state is therefore

\[
|\omega\rangle
=
\frac{1}{\sqrt{2}}
\left(
e^{i2\pi(0.617596)}|0\rangle
+
e^{i2\pi(0.132870)}|1\rangle
\right),
\]

\noindent where the relative phase between \(|0\rangle\) and \(|1\rangle\) is \(\Delta\phi = -2\pi(0.484726)\). The SHA-256 values may be reproduced using standard implementations, such as the online calculator provided in Ref. \cite{emn178_sha256}, which follows the NIST Secure Hash Standard (FIPS 180-4) \cite{nist_fips1804}.}

The cryptographically encoded QTM token can therefore be expressed as

\begin{equation}
\label{crypto_token}
\begin{aligned}
|\psi\rangle
&=
\frac{U_k}{\sqrt{2^q}}
\Biggl[
\bigotimes_{j=1}^{p}
U_{b_j}|u_j\rangle
\otimes
\sum_{x\in\{0,1\}^{q}}
e^{i2\pi H(s,x)}
|x\rangle \\
&\qquad\otimes
|0\rangle^{\otimes r}
\Biggr],
\end{aligned}
\end{equation}

\noindent prepared prior to each transaction.

Before proceeding to these phases, the QTM card may first be subjected to an identification check. Upon insertion, the CDD and a measured physical signature $F_t$ (e.g., Raman spectrum $R(\omega)$, defect-pattern image $I(x,y)$, photoluminescence spectrum $P(\lambda)$, etc. \cite{solin1970raman,Rohmfeld2000,Gruber1997,castelletto2014silicon}) can be sent to the bank for duplicate detection, thereby verifying that the presented card corresponds to a legitimate bank-issued card.

The signature $F_t \in \left\{ R(\omega),\, I(x,y),\, P(\lambda) \right\}$ derived from the characterization is associated with the intrinsic structural and spectroscopic properties of the underlying material. In particular, diamond and silicon carbide are structurally robust materials whose lattice and defect configurations are expected to evolve only slowly under normal operating conditions. Consequently, although the measured signatures may not remain numerically identical over long periods of use, they are expected to remain sufficiently distinguishable for similarity-based identification. Let $F_0$ denote the enrolled physical signature. The card is accepted only if $S(F_t,F_0)\ge\tau$, where $S(\cdot,\cdot)$ is a similarity function and $\tau$ is a predefined acceptance threshold. Thus $(\mathrm{CCD},F_t)
\xrightarrow{\;S(F_t,F_0)\ge\tau\;}|\psi\rangle$.

\section{Physical Realization}
\label{sec:Physical Realization}

This section presents an illustrative physical implementation of the cryptographic token introduced in Eq.~\eqref{crypto_token}. For simplicity, let $(p,q,r)=(1,1,1)$ be selected by the bank, together with $U_b=H$ and $U_k = I$, where $q_1\in\mathcal{S}$, $q_2\in\mathcal{R}$, and $q_3\in\mathcal{A}$. 

\begin{figure}
    \centering
\includegraphics[width=0.45\textwidth]{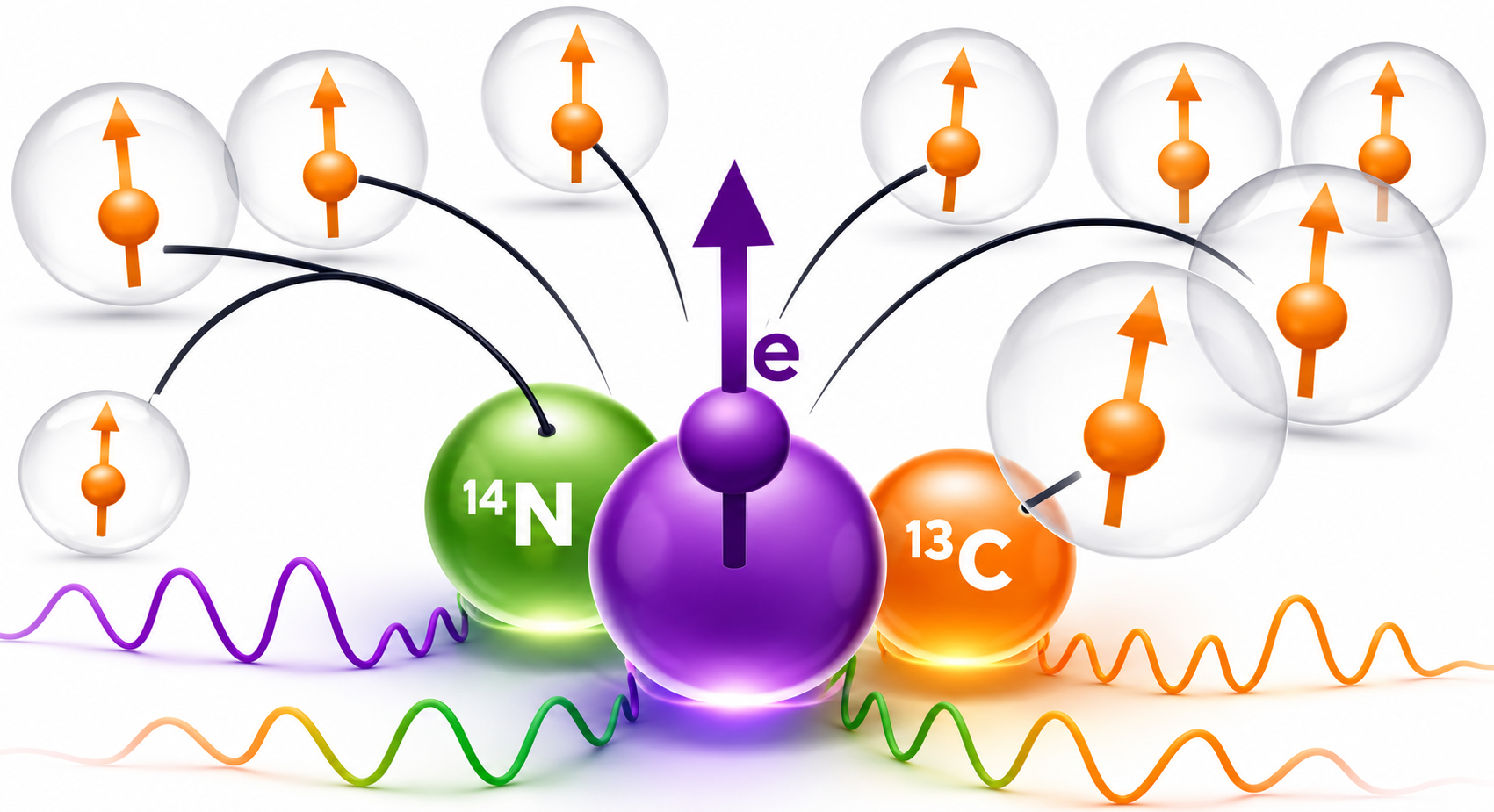}
\vspace{0.1 em} % Adjust the negative value as needed to reduce the gap
\caption{Conceptual representation of the NV center spin system, showing the electron spin, the host $^{14}\mathrm{N}$ nuclear spin, and the surrounding $^{13}\mathrm{C}$ spins. Inspired by Ref.~\cite{bradley2019ten}. The total Hilbert space is expressed as
$\mathcal{H}_{\mathrm{tot}}=\mathcal{H}_{e}\otimes\mathcal{H}_{^{14}\mathrm{N}}\otimes\bigotimes_{k>1}^{N}\mathcal{H}_{^{13}\mathrm{C},k}$,
where the surrounding $^{13}\mathrm{C}$ nuclei constitute the local spin environment of the NV center embedded in the quantum chip shown in Fig.~\ref{fig:1}.}
\label{fig:3}
\end{figure}

Assuming an NV-center in diamond implementation for the case, let \(q_1 \equiv e,
q_2 \equiv {}^{13}\mathrm{C},
q_3 \equiv {}^{14}\mathrm{N}\), where $e$ denotes the NV electron-spin qubit, while ${}^{13}\mathrm{C}$ and ${}^{14}\mathrm{N}$ denote the carbon and nitrogen nuclear-spin qubits, respectively. Figure~\ref{fig:3} shows the corresponding NV-center spin system, where $q_2$ and $q_3$ are coupled to the $q_1$.

(i)\(\textit{QTM actions for $e$ qubit}\). Before a token preparation on the QTM card, the electron spin is initialized in the \(|0\rangle_e \equiv |m_s = 0\rangle\) state by optical pumping using a \(532~\mathrm{nm}\) laser. Physically, the bank demand $U_b = H$ can be implemented by applying a microwave (MW) $\pi/2$ pulse through the terminal to drive transitions between the electron-spin sublevels, typically $|0\rangle_e \equiv |m_s = 0\rangle$ and $|1\rangle_e \equiv |m_s = -1\rangle$. 

The transition frequencies are determined by the zero-field splitting $D \approx 2.87~\mathrm{GHz}$ between the $m_s = 0$ sublevel and the degenerate $m_s = \pm 1$ sublevels \cite{Arya2025NV}. Under an applied magnetic field $B_0$, the $m_s = \pm 1$ levels undergo Zeeman splitting, yielding transition frequencies $f_{\pm}=D\pm\gamma_e B_0$, where $\gamma_e/2\pi \approx 28.025~\mathrm{MHz/mT}$ is the electron gyromagnetic ratio \cite{Abe2023}.

Coherent manipulation is performed by a magnetic field $B_{\mathrm{MW}}(t)=B_1\cos(\Omega_{\mathrm{MW}}t)$, where the microwave frequency $\Omega_{\mathrm{MW}}/2\pi=f_{\pm}$ is tuned to the targeted frequency, resulting Rabi frequency $\Omega_R=\gamma_e B_1$ \cite{Dolde2014NV, jelezko2004observation}. This enables selective addressing of the
$|m_s=0\rangle\leftrightarrow|m_s=-1\rangle$ transition, thereby preparing the superposition state $|m_s=0\rangle_e
\xrightarrow{\pi/2\ \mathrm{MW}} |u\rangle_e$, where

\begin{equation}
|u\rangle_e =
\frac{|m_s=0\rangle_e+|m_s=-1\rangle_e}{\sqrt2}.   
\end{equation}

(ii)\(\textit{QTM actions for ${}^{13}\mathrm{C}$ qubit}\). Next, the cryptographic data $(\theta_0,\theta_1)$ specified by the bank are encoded in a nearby ${}^{13}\mathrm{C}$ qubit, whose spin-$\frac{1}{2}$ degree of freedom binds the token state to transaction-specific parameters. The logical basis states are defined as $|0\rangle_{{}^{13}\mathrm C}\equiv\left|m_I=\frac{1}{2}\right\rangle$, and $|1\rangle_{{}^{13}\mathrm C}\equiv\left|m_I=-\frac{1}{2}\right\rangle$. 

When the QTM terminal prepares the electron in the $|m_s=0\rangle_e$ state, the nearby ${}^{13}\mathrm{C}$ nucleus experiences negligible hyperfine-induced frequency shifts and therefore evolves under the same external $B_0$ at the Larmor frequency, $f_{C_L}=\frac{\gamma_C B_0}{2\pi}$, where $\gamma_C/2\pi\approx10.705~\mathrm{MHz/T}$ is the ${}^{13}\mathrm{C}$ gyromagnetic ratio. In the $|m_s=-1\rangle_e$ manifold, the hyperfine interaction due to the local magnetic environment results in a shifted precession frequency \cite{childress2006coherent,dutt2007quantum}.
 
Prior to encoding, the ${}^{13}\mathrm{C}$ nucleus is initialized to its ground state $\left|m_I=\frac{1}{2}\right\rangle$ by repeatedly repolarizing the electron spin into the $|m_s=0\rangle_e$ state using $532~\mathrm{nm}$ optical pulses, thereby increasing the nuclear-spin polarization \cite{dutt2007quantum}. A resonant RF $\pi/2$-pulse then prepares the superposition state 

\begin{equation}
\label{carbon_nuclear_spin}
 \left|m_I=\frac{1}{2}\right\rangle
\xrightarrow{\pi/2\ \mathrm{RF}}
\frac{
\left|m_I=\frac{1}{2}\right\rangle+
\left|m_I=-\frac{1}{2}\right\rangle
}{\sqrt{2}}.   
\end{equation}

Subsequent encoding with the relative phase $\Delta$, yields the state as

\begin{equation}
\label{omega}
 |\omega\rangle_{{}^{13}\mathrm C} = \frac{1}{\sqrt{2}} \left(\left|m_I=\frac{1}{2}\right\rangle + e^{i\Delta} \left|m_I=-\frac{1}{2}\right\rangle \right).
\end{equation}

One way to implement phase encoding is through a virtual Z-gate, which corresponds to adding a phase offset to the drive field \cite{mckay2017efficient}. Operationally, this frame update can transform Eq.\eqref{carbon_nuclear_spin} to Eq.\eqref{omega}, thus imprinting the bank data phase in to the $q_2$ qubit.

(iii)\(\textit{QTM actions for ${}^{14}\mathrm{N}$ qubit}\). Following the actions of $q_1$ and $q_2$, the actions of $q_3$ due to nuclear spin $I=1$ can be defined in its hyperfine states $m_I\in\{-1,0,1\}$. The logical basis states are then defined as $|0\rangle_{{}^{14}\mathrm N}\equiv\left|m_I=0\right\rangle$, and $|1\rangle_{{}^{14}\mathrm N}\equiv\left|m_I=-1\right\rangle$. 

Concerning token preparation, the first step that follows are $q_2$ and $q_3$ initialization. This approach can be motivated by the ${}^{15}\mathrm{N}$--${}^{13}\mathrm{C}$ scheme demonstrated in Ref.~\cite{jacques2009dynamic}, where, near the excited-state level anticrossing, hyperfine-assisted polarization transfer enables simultaneous polarization of the ${}^{15}\mathrm{N}$ and the ${}^{13}\mathrm{C}$ spins, thereby realizing deterministic initialization of a three-qubit register. By analogy, the same mechanism may be extended to the $q_2$--$q_3$ system considered here.

Therefore, a token prepared by the terminal under the actions specified by the bank can be expressed as

\begin{widetext}
\begin{equation}
\begin{aligned}
|\Psi\rangle
=
\frac12\Big[&|m_s=0\rangle|m_I=-\tfrac12\rangle|m_I=0\rangle + e^{i\Delta}|m_s=0\rangle|m_I= \tfrac12\rangle|m_I=0\rangle\\&
+|m_s=-1\rangle|m_I=-\tfrac12\rangle|m_I=0\rangle + e^{i\Delta}
|m_s=-1\rangle
|m_I= \tfrac12\rangle
|m_I=0\rangle
\Big],
\end{aligned}
\end{equation}
\end{widetext}

\noindent which serves as a proof-of-concept implementation for the considered three-qubit register, while establishing a foundation for scalable multi-qubit realizations with extended and enhanced security. Such realizations may incorporate suitable entangling gates to generate entangled token states, providing an additional layer of authentication and further strengthening the overall security of the proposed QTM framework.

For verification, the bank can perform projective measurements in the prescribed bases and compare the observed outcomes with the expected responses. Agreement within an acceptable tolerance authenticates the token, whereas deviations result in rejection. Alternative quantum measurement and verification strategies may also be employed, depending on the underlying hardware platform and implementation requirements.

%One possible approach is for the bank to perform projective measurements, in which the observed results are compared with the expected statistics. Agreement within an acceptable tolerance authenticates the token, whereas any statistically significant deviation results in rejection. The implementation can be validated through quantum circuit simulations, which enable verification of the state-preparation and authentication procedures under both ideal and noisy conditions prior to experimental realization.

\section{Conclusions and outlook}
\label{sec:Conclusions and outlook}

This work presented a Quantum Teller Machine (QTM)-based framework for secure cash withdrawals using quantum-token authentication. The proposed NV-center implementation suggests that such a framework is compatible with experimentally demonstrated room-temperature quantum control techniques. Although practical deployment remains challenging, continued advances in solid-state quantum hardware are expected to facilitate its realization. Future work will focus on experimental implementation, protocol optimization, and the development of scalable quantum banking infrastructures. This may inspire the practical development of next-generation quantum-secure financial systems.

\bibliography{apssamp,references}% Produces the bibliography via BibTeX.

\end{document}